\documentclass[sigconf]{acmart}
\AtBeginDocument{%
  }

\copyrightyear{2026}
\acmYear{2026}
\setcopyright{cc}
\setcctype{by-nc-nd}
\acmConference[CHI '26]{Proceedings of the 2026 CHI Conference on Human Factors in Computing Systems}{April 13--17, 2026}{Barcelona, Spain}
\acmBooktitle{Proceedings of the 2026 CHI Conference on Human Factors in Computing Systems (CHI '26), April 13--17, 2026, Barcelona, Spain}
\acmDOI{10.1145/3772318.3790397}
\acmISBN{979-8-4007-2278-3/2026/04}

\usepackage{stfloats} 
\usepackage{tabularx}
\usepackage{cleveref}
\usepackage{multirow}
\usepackage{subfig}
\usepackage{xcolor}

\definecolor{nofeed}{HTML}{3C78D8}
\definecolor{control}{HTML}{6AA84F}
\definecolor{humanoid}{HTML}{E69138}
\definecolor{human}{HTML}{674EA7}

\begin{document}

\title{HumanoidTurk: Expanding VR Haptics with Humanoids for Driving Simulations}

\author{DaeHo Lee}
\orcid{0009-0007-9149-5676}
\affiliation{
  \institution{Department of AI Convergence, Gwangju Institute of Science and Technology}
  \city{Gwangju}
  \country{Republic of Korea}
}
\email{leedaeho@gm.gist.ac.kr}

\author{Ryo Suzuki}
\orcid{0000-0003-3294-9555}
\authornote{Co-Corresponding Authors}
\affiliation{
  \institution{University of Colorado Boulder}
  \city{Boulder}
  \country{USA}
}
\email{ryo.suzuki@colorado.edu}

\author{Jin-Hyuk Hong}
\orcid{0000-0002-8838-5667}
\authornotemark[1]
\affiliation{%
  \institution{Department of AI Convergence, Gwangju Institute of Science and Technology}
  \city{Gwangju}
  \country{Republic of Korea}
  }
\email{jh7.hong@gist.ac.kr}


\begin{abstract}
We explore how humanoid robots can be repurposed as haptic media, extending beyond their conventional role as social, assistive, collaborative agents. To illustrate this approach, we implemented HumanoidTurk, taking a first step toward a humanoid-based haptic system that translates in-game g-force signals into synchronized motion feedback in VR driving. A pilot study involving six participants compared two synthesis methods, leading us to adopt a filter-based approach for smoother and more realistic feedback. A subsequent study with sixteen participants evaluated four conditions: no-feedback, controller, humanoid+controller, and human+controller. Results showed that humanoid feedback enhanced immersion, realism, and enjoyment, while introducing moderate costs in terms of comfort and simulation sickness. Interviews further highlighted the robot’s consistency and predictability in contrast to the adaptability of human feedback. From these findings, we identify fidelity, adaptability, and versatility as emerging themes, positioning humanoids as a distinct haptic modality for immersive VR.
\end{abstract}



\begin{teaserfigure}
\centering
\includegraphics[width=\textwidth]{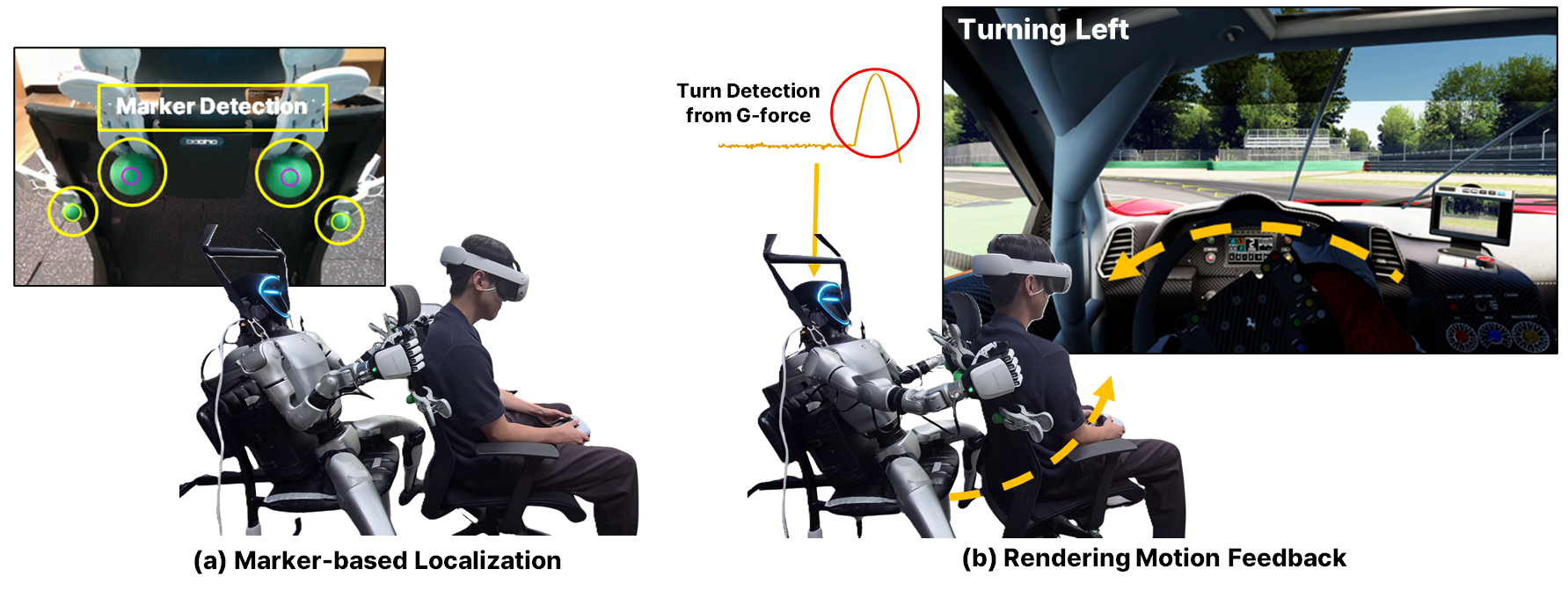}
\caption{We introduce HumanoidTurk, a humanoid robot repurposed to provide whole-body haptic feedback in VR driving. (a) The robot localizes and grasps the chair using marker-based detection. (b) In-game g-force signals are mapped into synchronized chair motion feedback, synthesizing immersive motion feedback to the VR user.}
\label{fig:teaser}
\Description[Teaser figure of HumanoidTurk]{A composite figure illustrating the HumanoidTurk system. On the left, AR markers attached to a chair are highlighted for marker-based localization. In the center, a humanoid robot grasps the chair while a VR user is seated, showing turn detection from g-force signals. On the right, a driving simulator screen shows a car turning left, with arrows indicating motion, and the robot delivering synchronized chair movement feedback to the user.}
\label{fig:teaser}
\end{teaserfigure}

\maketitle

\section{Introduction}
Haptics play a critical role in making VR experiences realistic and immersive. Prior research has explored diverse forms of haptic feedback, including vibration, force, and kinesthetic motion, each contributing to touch, manipulation, and movement feedback \citep{huang2020haptic, lecuyer2004can, dangxiao2019haptic}. In addition to modality, haptic systems also vary in scale, ranging from small wearable devices that stimulate localized body parts to large robotic or motion-platform systems that engage the whole body.

This distinction becomes most evident in the sensation of self-motion. Large-scale motion platforms deliver high-fidelity whole-body cues but remain bulky and fixed installations \citep{drivingvibe, hancock2008human}. In contrast, controller vibrations and wearables provide fine-grained yet highly localized cues, making them accessible but limited in fidelity \citep{hapseat, costes2022kinesthetic, costes2023inducing}. In addition to these limitations, most existing haptic systems, regardless of size, are purpose-built for a single function, offering limited versatility. To address this gap, we turn to humanoid robots, which combine mobility, multi-DOF manipulation, and human-like embodiment, suggesting a pathway to repurposable, human-scale physical feedback. Rather than designing a dedicated motion platform or a pair of industrial robot arms, we deliberately focus on general-purpose humanoids. In human-centric environments such as living rooms, labs, or kitchens, humanoids are increasingly explored as social or assistive agents \citep{iwamura2011elderly, mahdi2022survey, avataroid, zhao2006humanoid}. Repurposing the same robot as an interactive medium of haptic feedback allows a single platform to flexibly switch roles, acting as a companion or assistant and as a motion-feedback device alternatively, without additional fixtures or large-scale installations.

Building on this perspective, we present HumanoidTurk, taking the first step toward using humanoid robots as a medium for VR motion feedback. Our implementation translates in-game g-force features from a driving simulator into coordinated arm motions that move the user’s chair, producing synchronized whole-body cues. By physically engaging the body, the humanoid complements conventional motion feedback with embodied, human-scale feedback.

We evaluated this implementation through a two-phase study. A pilot study (n=6) compared two synthesis methods for mapping g-force to robot motion, from which we refined and selected a filter-based approach for smoother and more realistic dynamics. A main study (n=16) then evaluated humanoid feedback against three alternatives: no feedback, controller vibration, and human-delivered motion feedback. Results showed that humanoid feedback improved immersion and presence relative to no feedback and controller vibration, with a modest increase in discomfort and fatigue. Interview data further underscored its consistency and synchrony, while contrasting these strengths with the subtle adaptability unique to human-delivered feedback.

Our contributions are as follows:

(1) Repurposing a general-purpose humanoid to deliver embodied, whole-body motion feedback in VR

(2) Empirical evidence from a two-phase user study showing improvements in immersion and presence compared to no feedback and controller vibration, and presents meaningful insights by the comparison with human-delivered motion feedback.

(3) Suggesting emerging themes framing humanoid-mediated haptics along fidelity, adaptability, and versatility, outlining opportunities and challenges for future VR haptic systems.

\section{Related Work}
\subsection{Haptic Feedback in VR}
Haptics in VR encompasses diverse forms of feedback, including tactile (vibrotactile), thermal, and kinesthetic (motion and force feedback). To deliver these sensations, prior research has employed a wide range of media, which can be broadly categorized by their scale: large-scale (room-scale) and small-scale approaches.

Large-scale systems typically employ robotic or motion-based setups. For instance, robots have been used to reposition or morph objects to provide room-scale tactile experiences \citep{suzuki2020roomshift, gomi2024inflatablebots, weng2025hit}. Ikeda et al. mounted haptic properties onto a wheeled robot to provide continuous grounded feedback in a room-scale environment. For kinesthetic feedback, motion platforms remain the most common method, widely studied and commercialized in contexts such as 4D theaters \citep{YawVR3, QS210, danieau2014toward, stewart1965platform, vampire}. Creative variations even recruited people as actuators, as in Lung-Pan et al.’s HapticTurk, where multiple participants physically moved a user’s body \citep{hapticturk}. While such large-scale methods can deliver high-fidelity whole-body feedback, they are typically bulky and require substantial physical space \citep{drivingvibe, bouyer2017inducing}.

In contrast, small-scale systems focus on wearable or hand-held devices \citep{kim2024big, zenner2019drag, lee2019torc}. These include both invasive forms that attach directly to the body \citep{liu2025pneutouch, batik2025shiftly, jung2024hapmotion} and non-invasive devices such as hand-held controllers with haptic actuation \citep{tanaka2024haptic, hwang2024ergopulse}. Thermal feedback is generally implemented through wearables that integrate heating and cooling elements to simulate environmental temperature \citep{brooks2020trigeminal, gunther2020therminator, kang2024flip}. For kinesthetic feedback, researchers have explored compact solutions such as vibrotactile gloves, clothing-pulling systems, and low-cost chair-based devices that approximate motion sensations \citep{hoppe2021odin, costes2022kinesthetic, liu2020headblaster, hapseat, oishi2016enhancement}.

Taken together, these studies highlight the richness yet fragmentation of VR haptics. Most approaches remain narrowly scoped, focusing on specific body parts or single-purpose functions. This motivates our exploration of humanoid robots as a versatile modality for delivering embodied, whole-body feedback, extending VR haptics beyond localized or task-specific devices.

\subsection{Robots as Interactive Media}
Robots have long been explored not only as tools but also as interactive media that mediate between the physical environment and user experience. For instance, Shapebots demonstrated how shape-changing swarm robots can provide dynamic physical interaction in everyday spaces \citep{suzuki2019shapebots}. At the same time, MoveVR leveraged a household cleaning robot to deliver multiform force feedback in VR \citep{wang2020movevr}. Recent work has further emphasized how robots can be reappropriated beyond their primary functions. For example, Shiokawa et al. systematically investigated how the idle time of domestic robots could be leveraged to support novel ubiquitous interactions, introducing a multi-dimensional design space and proof-of-concept prototypes \citep{shiokawa2025beyond}. Such studies highlight that robots, even when designed for mundane chores, can be reframed as flexible platforms for broader interaction.

Among robots, humanoids have been studied most extensively due to their human-like form and movements. Early research positioned humanoids as cooperative partners \citep{breazeal2004humanoid, chang2010exploring}, companions \citep{ahmed2024human}, and learners of social cues \citep{calinon2006teaching}. Studies have also investigated how embodiment and presence shape social telepresence when mediated through humanoid forms \citep{kawaguchi2016effect, bainbridge2008effect}. These works collectively highlight the humanoid not merely as a machine that executes commands but as a social partner capable of forming relationships with people.

Yet, prior studies have primarily framed humanoids as social agents or assistive collaborators, with little attention to their potential as haptic media. Whereas most VR haptic systems are single-purpose devices, humanoids are inherently versatile: the same robot can act as a social partner, a physical assistant, or a provider of embodied haptic feedback. This work takes an early exploration that direction by exploring how humanoids can be repurposed as haptic media to deliver human-scale feedback in VR.

\section{Method}

\begin{figure*}[hb]
\centering
\includegraphics[width=\textwidth]{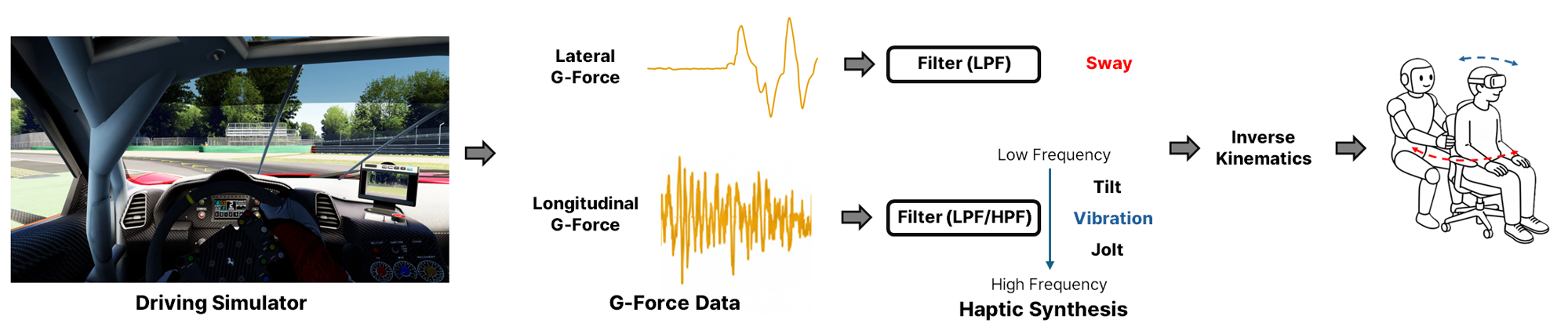}
\caption{Pipeline of our smoothed synthesis approach. In this example, when the participant drives over a curb while cornering, the LPF of lateral g-force generates a sway feedback. In contrast, the combination of LPF and HPF of longitudinal g-force triggers vibration feedback.}
\Description[A pipeline diagram]{Pipeline diagram showing how in-game driving forces are translated into humanoid-mediated haptic feedback. On the left, a racing simulator generates lateral and longitudinal g-force signals. These signals are filtered: lateral forces pass through a low-pass filter to produce chair sway, while longitudinal forces are split into low- and high-frequency components to generate tilt, vibration, and jolts. The resulting haptic cues are then mapped through inverse kinematics to control the humanoid robot, which grasps and moves the user’s chair to deliver synchronized whole-body feedback.}
\label{fig:experimentsetup}
\end{figure*}

\subsection{Apparatus}\label{sec:Apparatus}
We implemented HumanoidTurk using the Unitree G1 humanoid robot \citep{G1} equipped with Inspire Robots’ RH56DFTP hands \citep{RH56DFTP}. Each arm provides 9 degrees of freedom with a maximum payload of 3 kg, while each hand delivers up to 30 N of grasping force, allowing the humanoid to perform diverse arm motions and grasp different types of chairs. For the simulation environment, we used Assetto Corsa, a widely adopted driving simulator in prior motion-feedback research \citep{drivingvibe, liu2020headblaster, lee2024telemetry}. The game provides inertia values, including longitudinal and lateral g-forces, through both SharedMemory and UDP interfaces. We adopted the SharedMemory method for lower latency, receiving data at ~60–65 Hz. The VR environment was run on a high-performance desktop PC (Intel i9 CPU, RTX 3090 GPU, 48 GB RAM) connected via Quest Link to a Meta Quest 3 headset, running at 72 Hz. The task was performed on the Monza circuit and a Ferrari 458 GT2 in \textit{gamer} mode, which enables anti-lock braking and traction control systems but no other driving assists or AI-controlled cars. Participants controlled the car using a DualSense controller; the left stick for steering and the triggers for throttle and braking. We used only the controller's vibration, disabling adaptive triggers.

To support different chair types without additional hardware modifications, we developed a marker-based localization system. Two spherical markers per arm were tracked using the G1’s integrated RealSense D435i depth camera, allowing the robot to estimate its 3D positions relative to the body origin. We chose this simple marker-based approach instead of fully marker-free chair perception because inferring graspable regions on arbitrary chairs in unstructured real-world environments would require complex probabilistic computer vision pipelines that fall outside the scope of this work. This enabled the humanoid to grasp chairs and apply motion feedback to them without structural adjustments, making the system more adaptable to various types of chairs.

\subsection{Safety Considerations}
Humanoid robots can pose risks in the event of unexpected failures. We implemented multiple safety measures to minimize potential hazards. First, the humanoid was kept in a seated posture throughout experiments to reduce the chance of tipping or falling. Second, while markers provided chair localization, the final grasping motion was manually fine-tuned by the experimenter before the study began. Third, the robot was physically secured with an additional stand, serving as a backup safety mechanism. Finally, we implemented a software-level emergency stop that could instantly halt robot motion in case of anomalies such as overheating or process errors.

\subsection{Haptic Feedback Synthesis}
To implement feedback synthesis, we captured real-time G-force and telemetry data via the game’s Shared Memory interface, mapping continuous signals rather than relying on discrete event detection (e.g., acceleration or cornering), ensuring fluid feedback. As part of demonstrating the system, we implemented and tested two approaches to map g-force into chair motion cues. This comparison was not our primary research focus, but rather a practical step to identify a synthesis approach suitable for subsequent evaluation.

\textbf{Smoothed (filter-based) synthesis}:
In this approach, each axis of in-game g-force is decomposed into a slow component (via low-pass filtering) and a fast component (via high-pass filtering or jerk). Slow trends are used to produce continuous posture changes, while fast changes generate brief impacts. Concretely, lateral g-force drives chair yaw (sway) through its low-pass component, and longitudinal g-force drives chair pitch (tilt) through its low-pass component, while its fast component produces jolt and vibration. When multiple cues occur simultaneously, they are dynamically blended.

\textbf{Threshold-based synthesis}:
In contrast, the threshold-based method avoids continuous decomposition. Lateral g-force is still synthesized through a low-pass sway signal, but longitudinal g-force is mapped in a simpler, event-driven fashion. Whenever its magnitude or derivative exceeds a threshold, the system triggers discrete motion feedback. This design favors responsiveness and simplicity over continuity, offering a more event-driven alternative to the smoothed approach.

For both synthesis approaches, safety limits are applied, including longitudinal displacement ($\pm$10 cm), rotation ($\pm$20°), and participant weight (80 kg), to prevent overheating and unexpected failure. Additional details, such as cutoff frequency and threshold, are provided in Appendix A. To support reproducibility and encourage follow-up work, we will release our full implementation as open-source artifacts through git.

\subsection{Pilot Study}\label{sec:pilot}
We conducted a two-phase pilot study to refine and compare the two synthesis approaches iteratively. In the first phase, three participants experienced replay sessions using both methods and provided feedback, which we incorporated into parameter adjustments through multiple refinement cycles. In the second phase, three additional participants compared the improved versions of both systems during live driving sessions. All three participants preferred the Smoothed synthesis approach, citing higher realism and better alignment with driving dynamics. In both phases, after all sessions, one of the researchers practiced the human-delivered feedback condition for the \textit{Human+Controller} condition. The performer was a male (186 cm, 94 kg) who was physically capable of exerting sufficient force on the chair occupied by the participants. He observed the shared game screen and listened to the sound to synchronize his actions with events such as braking and cornering. During this practice, participants reported the perceived intensity, timing, and type of feedback to establish a consistent way of delivering human feedback.

\begin{figure*}[!hb]
\centering
\includegraphics[width=\textwidth]{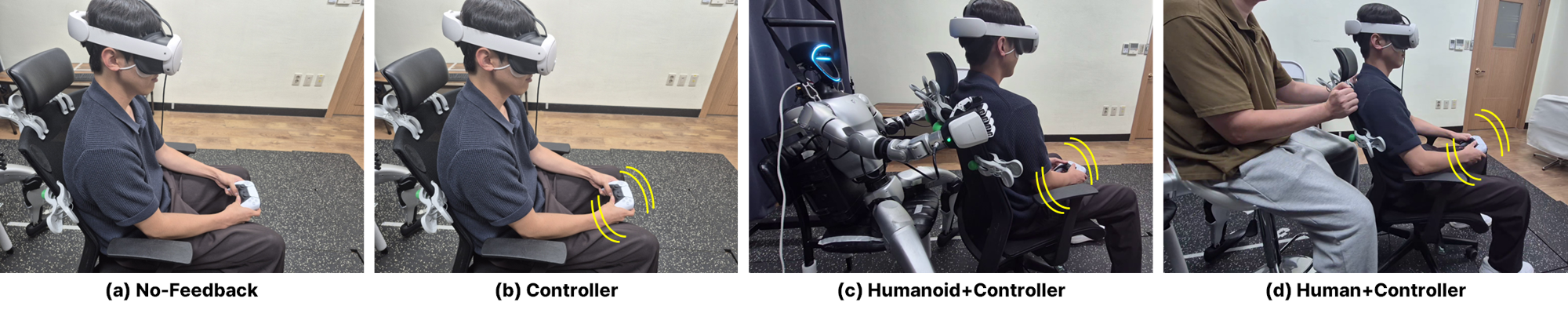}
\caption{Experimental setup for the four feedback conditions. (a) \textcolor{nofeed}{\textit{No-Feedback}}, where the participant drives in VR using only visual and auditory cues, and (b–d) conditions in which vibration is delivered through the game controller: (b) \textcolor{control}{\textit{Controller}}; (c) \textcolor{humanoid}{\textit{Humanoid+Controller}}, where the humanoid robot apply motion feedback to the chair; and (d) \textcolor{human}{\textit{Human+Controller}}, where a human operator behind the participant apply motion feedback to the chair.}
\Description[Three photos of experimental setup]{Three photos showing the experimental setup for evaluating humanoid-mediated haptic feedback in VR driving. Left: a participant wearing a VR headset sits on an office chair and holds a game controller. Center: the humanoid robot sits behind the participant, grasping the chair with both arms to deliver synchronized motion cues. Right: a human experimenter provides comparable chair perturbations by hand, serving as the human-control condition.}
\label{fig:experimentsetup}
\end{figure*}

\subsection{Technical Evaluation}
We conducted a technical evaluation to measure the delay of our system. Driving sessions were performed in Assetto Corsa on the Monza track with the Ferrari 458 GT2, where each session lasted 40-50 seconds (M=43.8). We measured total computational latency, including (1) delay from game to system G-force capture, (2) delay from G-force to haptic synthesis, (3) arm inverse kinematics computation time. Because latency can critically affect user experience and task accuracy in both VR and haptics \citep{jerald2015vr, louca2024impact}, we included these metrics to ensure that each stage of our pipeline was within acceptable bounds. Latency was recorded for every frame across five repeated sessions, yielding a total of 6,161 samples (approx. 1,100–1,400 frames per session). The overall computational latency was M = 34.7 ms, SD = 3.4. The majority of this delay stemmed from the inverse kinematics computation (M = 34.5 ms, SD = 3.4). These results confirm that the pipeline operates consistently under 40 ms, which is within acceptable limits for real-time haptic feedback \citep{frank1988effects}.

\section{User Study}
We conducted a user study with 16 participants to evaluate the effectiveness of HumanoidTurk. The experiment setup was identical to the pilot study, as described in Section~\ref{sec:Apparatus}. Participants experienced four conditions: (1) \textcolor{nofeed}{\textit{No-Feedback}}, (2) \textcolor{control}{\textit{Controller}}, (3) \textcolor{humanoid}{\textit{Humanoid+Controller}}, (4) \textcolor{human}{\textit{Human+Controller}}. After each condition, they completed questionnaires and a brief interview. The evaluation employed a combination of quantitative and qualitative methods, including the Simulation Sickness Questionnaire (SSQ), the User Experience Questionnaire–Short (UEQ-S), and additional self-authored items assessing immersion, presence, comfort, enjoyment, and suitability (see Appendix C), as well as follow-up interviews.

\subsection{Participants}
We recruited participants from a local university (6 females, 10 males; age M = 22.9 years, SD = 2.92). Twelve participants reported no or limited driving experience, two reported 1–2 years of driving experience, and two reported approximately 4 years of driving experience. For the VR experience, participants rated themselves on a 7-point Likert scale (M=3.3, SD=1.25, 1=Not experienced, 7=Regular user). Regarding prior simulation experiences (e.g., 4D movies, motion platforms), six participants reported no experience, four reported 1–2 times, six reported approximately five times, and one participant reported about ten times. Detailed demographics are summarized in the Appendix (Table~\ref{tab:demographics}).

\subsection{Procedure}
Upon arrival, participants were briefed on the study and asked to sign an informed consent form. They were informed that VR and motion feedback could cause discomfort or motion sickness and that they could withdraw at any time. After consent, demographic information was collected, including age, weight, driving experience, and VR experience.

Because the pilot study showed that novice drivers struggled to control the car in Assetto Corsa immediately, participants first practiced driving on a 2D monitor. Depending on their skill, participants practiced for 5–10 minutes until they demonstrated basic control. Next, they drove in VR without feedback to become accustomed to the VR environment and to establish baseline levels of motion sickness. Participants then completed the SSQ, and all were confirmed to be free of major discomfort before proceeding to the main sessions.

We used a Latin square design to counterbalance the order of conditions. Each session lasted one minute to limit fatigue and sickness, after which participants completed the SSQ, UEQ-S, and individual item questionnaires. The SSQ comprised 16 items on a 0–3 scale, the UEQ-S included eight items on a –3 to 3 scale, and the individual items (immersion, presence, comfort, enjoyment, suitability) were rated on a 7-point Likert scale. Immediately afterward, we conducted a short interview to capture participants’ impressions while their experiences were still fresh. These brief interviews focused on overall impressions and immediate reactions. The evaluation took about five minutes per session. Participants were encouraged to rest until any simulator sickness subsided. After completing all sessions, participants engaged in a final interview that probed opportunities and challenges for future applications of the humanoid haptic feedback system. The entire procedure took approximately 60 minutes, and participants received about 20,000 Won (about \$15) as compensation. This study was reviewed and approved by the Institutional Review Board (IRB).

Specifically, we included the \textcolor{nofeed}{\textit{No-Feedback}} condition to serve as a visual-only baseline. This establishes the fundamental level of simulation sickness caused by the visual-vestibular conflict inherent in VR driving, allowing us to isolate the effects of haptic mitigation. In the \textcolor{human}{\textit{Human+Controller}} condition, a trained researcher delivered motion feedback following the protocol established in the pilot study (Section~\ref{sec:pilot}), aiming to provide forces and timings that had been iteratively refined based on pilot participants’ feedback. While the \textcolor{human}{\textit{Human+Controller}} condition inevitably allowed for quantitative comparison with the other conditions, its primary purpose was to serve as a reference point for examining participants’ perceptual differences between human- and humanoid-mediated feedback. In other words, it functioned less as a baseline and more as a probe to contextualize how users interpret the humanoid’s feedback relative to a human’s adaptive feedback.

\section{Results}

\begin{figure}[!t]
\centering
\includegraphics[width=0.5\textwidth]{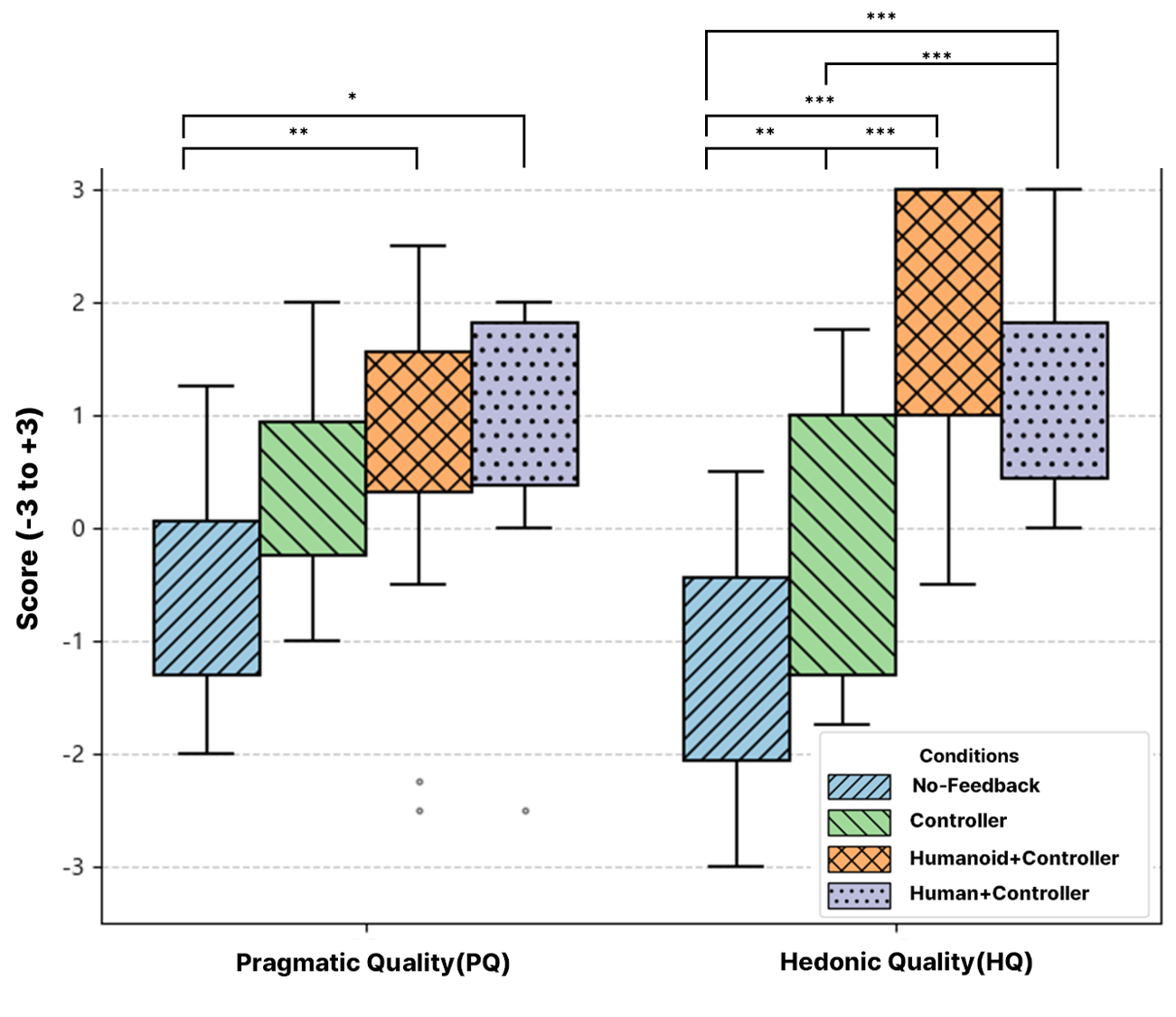}
\caption{
The boxplot of the results for Pragmatic Quality and Hedonic Quality from UEQ-S. The \textcolor{human}{\textit{Human+Controller}} condition achieved the highest PQ, while the \textcolor{humanoid}{\textit{Humanoid+Controller}} scored significantly higher in HQ, indicating superior enjoyment. The \textcolor{nofeed}{\textit{No-Feedback}} condition consistently scored lowest across both dimensions. (*\textit{p}<.05, **\textit{p}<.01, ***\textit{p}<.001)}
\Description{Boxplot showing user ratings of Pragmatic Quality (PQ) and Hedonic Quality (HQ) across four conditions: No-feedback (blue), Controller (orange), Controller+Humanoid (green), and Controller+Human (red). Scores range from -3 to +3. For PQ, No-feedback has the lowest median below zero, Controller is slightly above zero, while Controller+Humanoid and Controller+Human both have higher medians around +1. For HQ, No-feedback and Controller remain below zero, whereas Controller+Humanoid shows the highest median near +2, followed by Controller+Human around +1. Statistical significance markers indicate that Controller+Humanoid and Controller+Human are significantly higher than No-feedback and Controller.}
\label{fig:PQHQ}
\end{figure}

\begin{table*}[!ht]
\centering
\caption{Mean (and Standard Deviation) of SSQ Scores. The \textcolor{humanoid}{\textit{Humanoid+Controller}} condition resulted in the highest sickness scores across all subscales, showing an increase compared to the other conditions.}
\label{tab:ssq_scores}
\Description{Table summarizing Simulator Sickness Questionnaire (SSQ) results across four conditions. Each cell shows mean (standard deviation). For No-feedback: Nausea 4.77 (7.79), Oculomotor 10.42 (16.58), Disorientation 11.31 (22.28), Total Score 10.05 (16.14). For Controller: Nausea 4.77 (7.79), Oculomotor 7.58 (10.72), Disorientation 6.96 (12.45), Total Score 7.48 (11.09). For Controller + Humanoid: Nausea 17.89 (15.13), Oculomotor 18.95 (17.72), Disorientation 15.66 (21.49), Total Score 20.57 (18.42). For Controller + Human: Nausea 8.94 (8.86), Oculomotor 9.95 (11.32), Disorientation 7.83 (10.13), Total Score 10.52 (9.58).}
\begin{tabular}{lcccc}
\toprule
\textbf{Condition} & \textbf{Nausea} & \textbf{Oculomotor} & \textbf{Disorientation} & \textbf{Total Score} \\
\midrule
No-Feedback & 4.77 (7.79) & 10.42 (16.58) & 11.31 (22.28) & 10.05 (16.14) \\
Controller & 4.77 (7.79) & 7.58 (10.72) & 6.96 (12.45) & 7.48 (11.09) \\
Humanoid+Controller & 17.89 (15.13) & 18.95 (17.72) & 15.66 (21.49) & 20.57 (18.42) \\
Human+Controller & 8.94 (8.86) & 9.95 (11.32) & 7.83 (10.13) & 10.52 (9.58) \\
\bottomrule
\end{tabular}
\end{table*}

This section presents the results of the SSQ, UEQ-S, and individual items, along with illustrative quotes from interviews. Since our study employed a within-subjects design, we first tested for normality in all measures, confirming that none of the questionnaire results met the normality assumptions. We therefore used the non-parametric Friedman test, followed by Conover post-hoc tests with Holm correction.

\begin{figure*}[!b]
\centering
\includegraphics[width=0.7\textwidth]{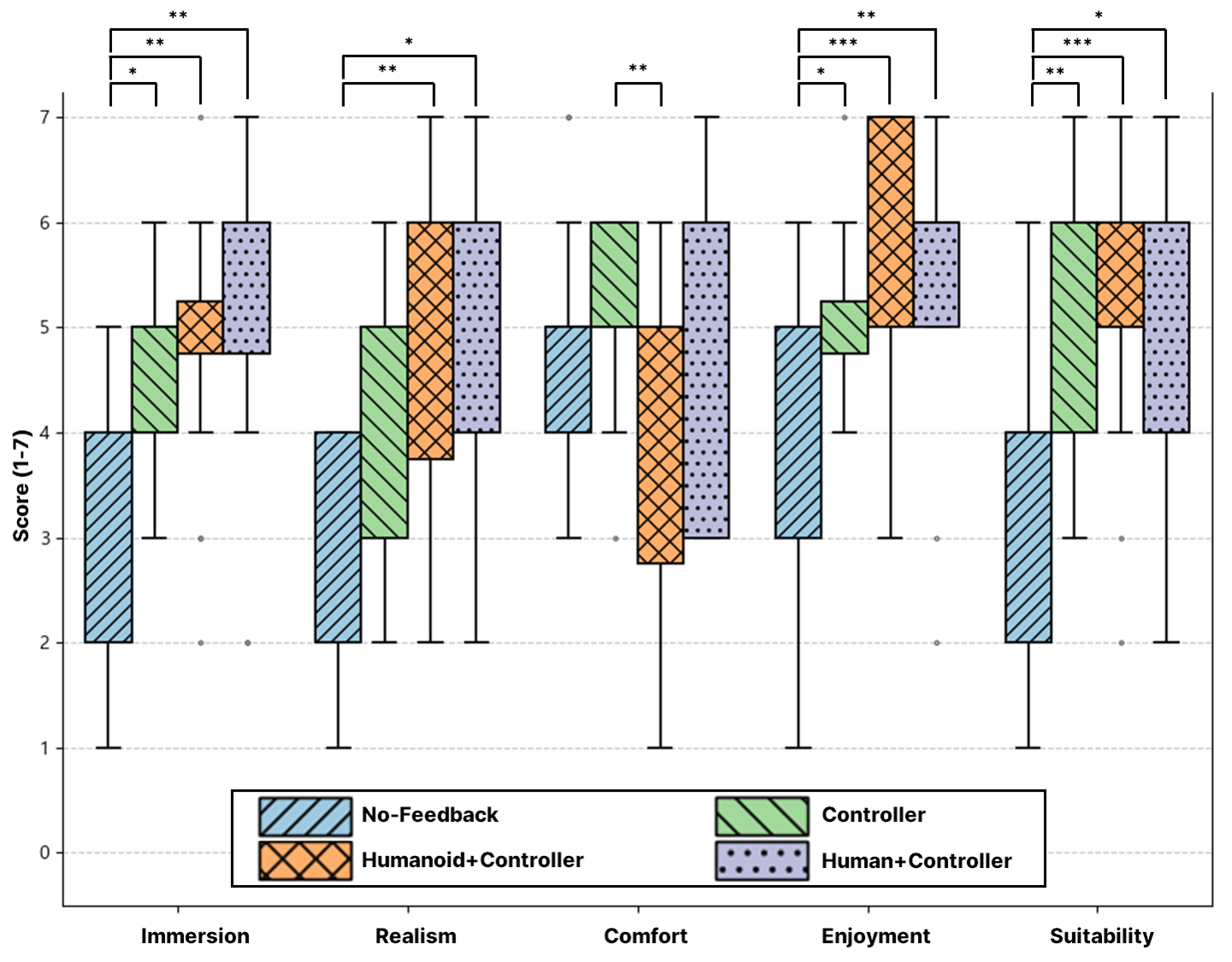}
\caption{
The boxplot of the results for individual questionnaires. The \textcolor{humanoid}{\textit{Humanoid+Controller}} and \textcolor{human}{\textit{Human+Controller}} conditions consistently achieved higher ratings for Immersion, Realism, Enjoyment, and Suitability compared to the baselines. However, a trade-off was observed in Comfort, where the \textcolor{control}{\textit{Controller}} condition was rated higher than the \textcolor{humanoid}{\textit{Humanoid+Controller}}. (*\textit{p}<.05, **\textit{p}<.01, ***\textit{p}<.001)}
\label{fig:indivQ}
\Description{Boxplot comparing participant ratings (1-7 scale) across four conditions: No-feedback (blue), Controller (orange), Controller+Humanoid (green), and Controller+Human (red). For Immersion, No-feedback has the lowest scores, while Controller+Humanoid and Controller+Human are significantly higher (**p<.01). For Realism, Controller+Humanoid and Controller+Human again score higher than No-feedback and Controller (*p<.05, **p<.01). For Comfort, Controller is rated highest, significantly above Controller+Humanoid (**p<.01). For Enjoyment, Controller+Humanoid is highest, followed by Controller+Human, both significantly above No-feedback and Controller (*p<.05, **p<.001). For Suitability, Controller+Humanoid and Controller+Human are significantly higher than No-feedback and Controller (**p<.01, p<.05). Overall, humanoid and human feedback conditions enhanced immersion, realism, enjoyment, and suitability, though comfort was lower for humanoid feedback.}
\end{figure*}

\subsection{User Experience (UEQ-S)}
UEQ-S results highlighted its positive experiential impact. Participants rated \textcolor{humanoid}{\textit{Humanoid+Controller}} highest on items such as Enjoyable, Exciting, and Interesting. Friedman tests showed significant effects for all items except Efficiency (\textit{p} = .081) and Clarity (\textit{p} = .204). When aggregated into the two standard dimensions, both pragmatic quality (\textit{p} < .01) and hedonic quality (\textit{p} < .001) showed significant differences across conditions.

Participants rated \textcolor{humanoid}{\textit{Humanoid+Controller}} highest on hedonic quality (HQ, M=1.88, SD=1.05), while \textcolor{human}{\textit{Human+Controller}} scored highest on pragmatic quality (PQ, M=0.84, SD=1.16). \textcolor{nofeed}{\textit{No-Feedback}} consistently scored lowest on both dimensions (see Fig~\ref{fig:PQHQ}). Post-hoc tests confirmed that \textcolor{nofeed}{\textit{No-Feedback}} was significantly lower than both \textcolor{humanoid}{\textit{Humanoid+Controller}} (\textit{p} < .01) and \textcolor{human}{\textit{Human+Controller}} (\textit{p} < .05) in pragmatic quality. In terms of hedonic quality, the \textcolor{nofeed}{\textit{No-Feedback}} and \textcolor{control}{\textit{Controller}} showed statistical significance compared to the \textcolor{humanoid}{\textit{Humanoid+Controller}} and \textcolor{human}{\textit{Human+Controller}} (\textit{p} < .001). To illustrate, three participants explicitly stated that humanoid vibrations were “fun” or “enjoyable,” while seven participants noted that the \textcolor{nofeed}{\textit{No-Feedback}} reduced the game’s enjoyment. Detailed analyses are provided in the Appendix (Table~\ref{tab:PQHQ}, Table~\ref{tab:ueqs_item_scores}).

\subsection{Simulation Sickness}
For the SSQ, weighted scores were calculated for the four subscales: Nausea, Oculomotor, Disorientation, and Total Score. Table~\ref{tab:ssq_scores} summarizes the mean and standard deviation values across conditions. Friedman tests revealed significant effects for Nausea (p < .001), Oculomotor (p < .01), and Total Score (p < .01), but not for Disorientation (p = .196). Among the four conditions, \textcolor{humanoid}{\textit{Humanoid+Controller}} consistently produced the highest scores.

Post-hoc analyses indicated that Nausea was significantly higher in \textcolor{humanoid}{\textit{Humanoid+Controller}} compared to \textcolor{nofeed}{\textit{No-Feedback}} (p < .001), \textcolor{control}{\textit{Controller}} (p < .001), and \textcolor{human}{\textit{Human+Controller}} (p < .01). For Oculomotor, \textcolor{humanoid}{\textit{Humanoid+Controller}} was significantly higher than \textcolor{control}{\textit{Controller}} (p < .001) and \textcolor{nofeed}{\textit{No-Feedback}} (p < .01). Total Scores were also significantly higher for \textcolor{humanoid}{\textit{Humanoid+Controller}} than for \textcolor{nofeed}{\textit{No-Feedback}} (p < .01) and \textcolor{control}{\textit{Controller}} (p < .001). No other pairwise comparisons reached significance.

Two participants reported simulation sickness in the \textcolor{humanoid}{\textit{Humanoid+}} \textcolor{humanoid}{\textit{Controller}} and \textcolor{human}{\textit{Human+Controller}} conditions, but all participants were able to complete the tasks. P2 described \textcolor{humanoid}{\textit{Humanoid+Controller}} as “\textit{a strong intensity and continuous micro-vibrations that made me feel simulation sickness from my body constantly moving.}” Regarding the \textcolor{human}{\textit{Human+Controller}}, P1 noted, “\textit{The constant movement of my body made me feel a bit motion sick},” while P2 added, “\textit{I felt a slight, car-sickness-like feeling, and the jolting vibration was the main cause of that discomfort.}” Detailed item-level analyses for each SSQ question are provided in the Appendix (Table~\ref{tab:ssq_all_items}).

\subsection{Individual Questionnaires}
We further examined participants’ ratings on immersion, presence, comfort, enjoyment, and suitability. As shown in Fig.~\ref{fig:indivQ}, \textcolor{humanoid}{\textit{Humanoid+Controller}} and \textcolor{human}{\textit{Human+Controller}} achieved higher ratings for most of the factors. In contrast, comfort was rated higher in the \textcolor{control}{\textit{Controller}} (M=5.06, SD=0.85) and \textcolor{nofeed}{\textit{No-Feedback}} (M=4.81, SD=1.17). Friedman tests revealed significant differences for all five measures.

Post-hoc analyses indicated that \textcolor{nofeed}{\textit{No-Feedback}} scored significantly lower than all other conditions for immersion and enjoyment (\textit{p} < .05). For presence, \textcolor{nofeed}{\textit{No-Feedback}} was significantly lower than \textcolor{humanoid}{\textit{Humanoid+Controller}} (\textit{p} < .01) and \textcolor{human}{\textit{Human+Controller}} (\textit{p} < .05). For comfort, \textcolor{control}{\textit{Controller}} was rated significantly higher than \textcolor{humanoid}{\textit{Humanoid+Controller}} (\textit{p} < .01). Interestingly, comfort ratings did not differ between \textcolor{nofeed}{\textit{No-Feedback}} and \textcolor{humanoid}{\textit{Humanoid+Controller}}, suggesting that both the absence of feedback and strong humanoid feedback may reduce comfort in different ways. For suitability, \textcolor{nofeed}{\textit{No-Feedback}} again scored significantly lower than all other conditions (\textit{p} < .05).

Interview data provided additional nuance. Most participants (N = 14) reported increased fatigue in the \textcolor{humanoid}{\textit{Humanoid+Controller}} condition. Common reasons included vibration intensity and duration. For example, P1 and P4 both remarked: “\textit{The vibration was so intense that it made me feel fatigued}.” Similarly, P3 commented: “\textit{I felt like it could get tiring if I did this for a long time.}” At the same time, 13 participants highlighted the timely nature of humanoid feedback, with repeated remarks such as “\textit{The timing felt right},” especially in comparison with the \textcolor{human}{\textit{Human+Controller}}. Overall, participants acknowledged a trade-off between comfort and enjoyment: while humanoid and human feedback enhanced immersion and fun, these benefits were sometimes offset by physical discomfort.

\subsection{Overall Preference}
\begin{figure}[t]
\centering
\includegraphics[width=0.48\textwidth]{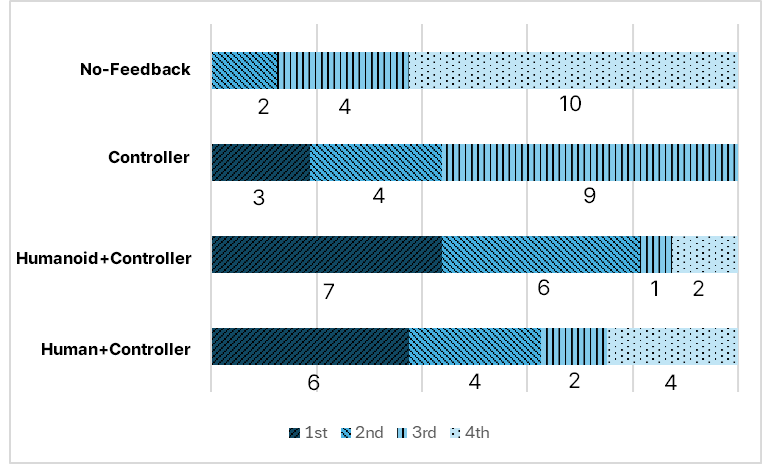}
\caption{
Overall Preference. Participants showed a strong preference for conditions with motion feedback, with the \textcolor{humanoid}{\textit{Humanoid+Controller}} and \textcolor{human}{\textit{Human+Controller}} conditions securing the majority of 1st and 2nd place rankings. In contrast, the \textcolor{nofeed}{\textit{No-Feedback}} condition was ranked last.}
\label{fig:preference}
\Description{Stacked bar chart showing participants’ overall preference rankings across four conditions: No-feedback, Controller, Controller+Humanoid, and Controller+Human. Lighter to darker shades represent 1st, 2nd, 3rd, and 4th place rankings. For No-feedback, most participants ranked it last (10 in 4th place, 4 in 3rd, 2 in 1st). For Controller, the majority also placed it last (9 in 4th, 4 in 3rd, 3 in 1st). For Controller+Humanoid, most ranked it highest (7 in 1st, 6 in 2nd, only 1 in 3rd, and 2 in 4th). For Controller+Human, responses were mixed but positive overall (6 in 1st, 4 in 2nd, 2 in 3rd, 4 in 4th). Overall, Controller+Humanoid and Controller+Human were the most preferred conditions, while No-feedback and Controller were mostly ranked lowest.}
\end{figure}

Fig.~\ref{fig:preference} summarizes participants’ overall preferences across the four conditions. Most participants favored either \textcolor{humanoid}{\textit{Humanoid+Controller}} (N=7) or \textcolor{human}{\textit{Human+Controller}} (N=6), with only three participants preferring the \textcolor{control}{\textit{Controller}}. Those who selected \textcolor{humanoid}{\textit{Humanoid+Controller}} often emphasized its realism and immersion. For instance, P6 remarked, “The humanoid moved the chair in sync with the vibrations, which made it feel much more realistic,” and P13 noted, “\textit{The immediate feedback in the right direction and intensity greatly enhanced immersion.}” Several participants also highlighted the novelty of humanoid feedback, such as P10, who stated, “It felt new to me since I haven’t experienced many simulators before,” and P11, who described it as “a new kind of feedback, which felt impressive.”

Participants who preferred \textcolor{human}{\textit{Human+Controller}} emphasized the smoothness of motion and the natural feel of acceleration/deceleration. P1 and P14 both commented on the “softness” of the feedback, which they found appealing. In contrast, two participants explicitly did not prefer \textcolor{humanoid}{\textit{Humanoid+Controller}}: P16 found the robot’s operational noise distracting, while P5 acknowledged its realism and enjoyment but remarked, “\textit{If I were to use it for a long time, comfort would matter more.}”

\section{Discussion}
Our findings highlight the differences between humanoid-mediated motion feedback and both conventional controller vibration and human-provided feedback. Beyond immediate comparisons, these results offer an opportunity to reflect more broadly on how humanoids and VR haptics should be designed to strike a balance between immersion, realism, and comfort. While our evaluation was limited to motion-based feedback, participants’ reflections revealed emerging themes that resonate with VR haptics more broadly.

\subsection{Emerging Themes for Humanoid-Mediated Haptics}
Our quantitative results and interviews reveal three recurring themes in how participants experienced the humanoid-based feedback: Fidelity, Adaptability, and Versatility.

\textbf{Fidelity} captures how precisely a system produces haptic feedback. Our results showed that \textcolor{humanoid}{\textit{Humanoid+Controller}} improved realism and enjoyment significantly compared to \textcolor{nofeed}{\textit{No-Feedback}} and \textcolor{control}{\textit{Controller}} (Section 5.1, 5.3). Participants explicitly described humanoid vibrations as “fun” or “enjoyable,” while others emphasized the enhanced realism of the moving chair (P6, P13). These findings suggest that humanoids can occupy a middle ground: delivering embodied, human-scale haptic feedback that enhances realism and immersion without requiring permanent infrastructure.

\textbf{Adaptability} describes the balance between mechanical consistency and adaptive responsiveness. In our study, participants appreciated the predictability and consistency of the humanoid feedback, repeatedly noting that “the timing felt right” (Section 5.3); however, they also criticized its lack of human-like subtlety compared to the \textcolor{human}{\textit{Human+Controller}} condition. This aligns with quantitative results, where \textcolor{human}{\textit{Human+Controller}} achieved the highest pragmatic quality (Section 5.1), suggesting that humans provide subtle adaptations that robots lack relatively. Humanoids make this trade-off more explicit, since their anthropomorphic form elicits expectations of social agency.

\textbf{Versatility} captures the degree to which a system can assume multiple roles beyond its immediate function as a haptic device. Unlike the other themes, this emerged mainly from participants' future-oriented comments rather than measured differences between conditions. Although we only evaluated a driving task, several participants discussed what else the robot could do in their immersive experiences. For example, P1 envisioned the robot holding a user’s hand during VR experiences to provide comfort or companionship, while P13 imagined playful interactions such as startling the user during a horror movie. These remarks highlight how participants saw the humanoid not only as a motion-feedback device but also as a flexible medium for everyday immersive experiences. We explore further opportunities of repurposing humanoid as haptic media in Section 6.2.2.

\begin{figure*}[!b]
\centering
\includegraphics[width=\textwidth]{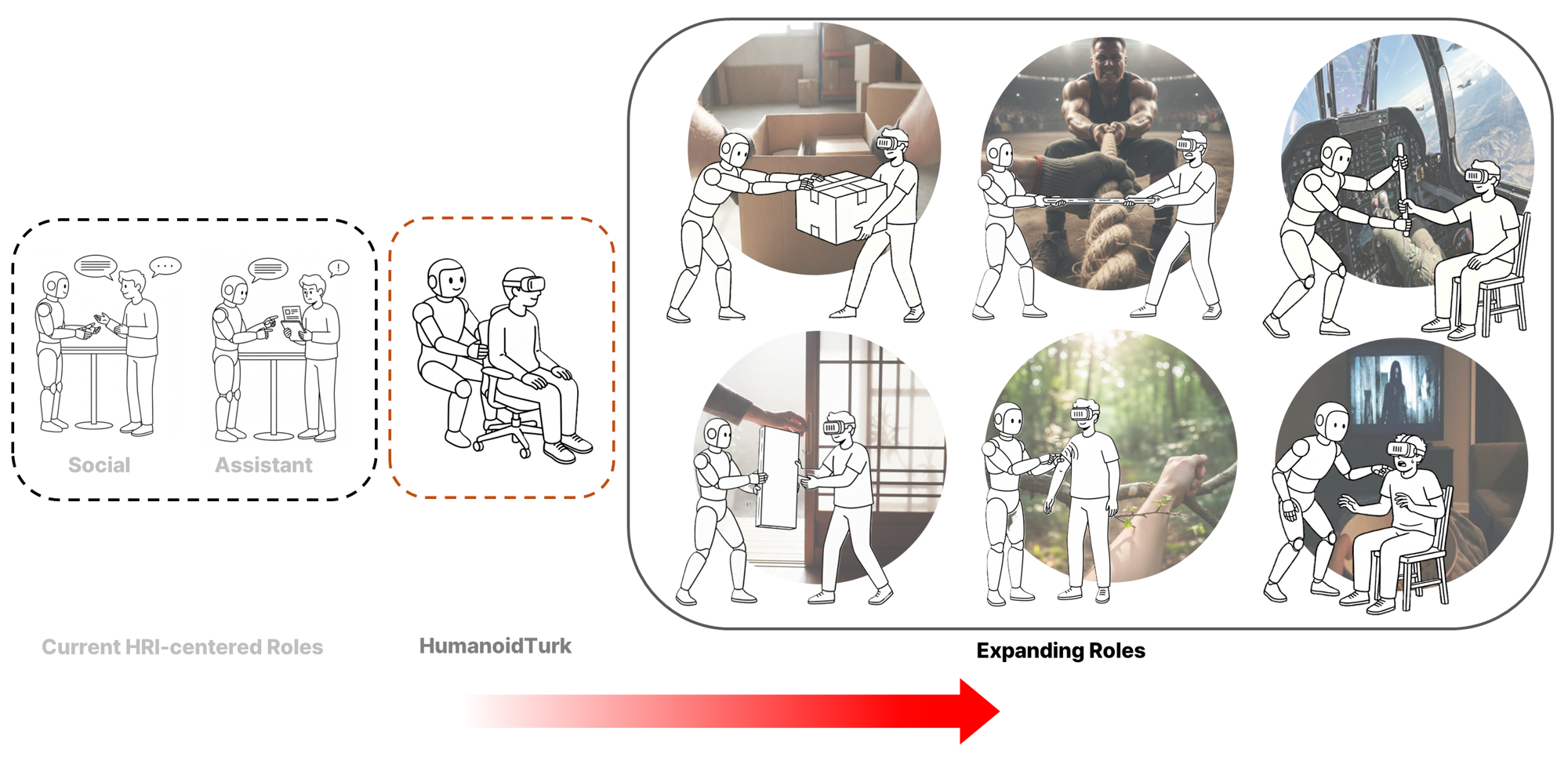}
\caption{Conceptual diagram of humanoid robots’ current HRI-centered roles (social, assistant), HumanoidTurk as a transitional step, and potential expanding roles. The 'Expanding Roles' section demonstrates versatility: (Top) rendering adjustable loads, providing resistance for physical training (e.g., tug-of-war), and modulating virtual joystick stiffness; (Bottom) providing haptic proxies for doors, rendering environmental textures (e.g., a leaf), and delivering active feedback for surprise effects. Example images were generated using the Gemini image generation model.}
\label{fig:discussion}
\Description{A conceptual diagram illustrating the evolution of humanoid robot roles, moving from current applications to future haptic possibilities. The diagram is divided into three stages from left to right: Current HRI-centered Roles: Depicts standard human-robot interaction scenarios where humanoids act as social companions or assistants. HumanoidTurk: Shows the system proposed in this paper as a transitional step, where a robot grasps a seated user's chair to provide synchronized motion feedback. Expanding Roles: Illustrates six diverse scenarios where humanoids function as versatile haptic media. These examples include: simulating adjustable weight by pushing a box; providing physical resistance for training exercises like tug-of-war; modulating the stiffness of a joystick in a vehicle simulation; acting as a physical proxy for opening a sliding door; rendering environmental textures, such as touching a leaf; and delivering active, surprise feedback by touching a user’s shoulder in a horror experience.}
\end{figure*}

\subsection{Current Position and Future Directions of Human-Humanoid Interactions}
In this section, we situate humanoid-mediated haptics within the current capabilities and limitations of off-the-shelf humanoid robots, and outline pathways for extending humanoid haptics to broader, more versatile interactive contexts.

\subsubsection{Current Position} Although humanoid robots have made substantial progress in mobility, dexterity, and autonomy, their physical reliability remains insufficient for sustained haptic feedback. In practice, we observed issues such as overheating and torque loss, which necessitated limiting task duration to one minute. For safety, we further constrained the robot to operate in a seated configuration rather than manipulating objects while standing. We also adopted a mediated interaction model, actuating the chair rather than the user directly. While direct interaction or standing manipulation could theoretically expand the haptic design space to include stronger forces and more diverse effects (e.g., simulating crashes), we prioritized mediated contact to ensure physical safety and mitigate user anxiety in this proof-of-concept study. These limitations underscore the existing gap between humanoid hardware and the requirements of sustained, high-fidelity physical interaction. As a result, humanoids today are more commonly positioned as assistive agents, collaborative partners, or social companions rather than reliable providers of embodied haptic feedback (see Sec.~6.1).

At the same time, humanoid robots exhibit multi-purpose capabilities enabled by their diverse sensing and software ecosystems. For instance, the Unitree G1 model we employed integrates two IMUs, a depth camera, and a LiDAR sensor, allowing it to perform environment perception without relying on external marker-based localization systems often required by dedicated room-scale haptic platforms. Moreover, the growing availability of open-source software and manufacturer-provided SDKs expands their utility; Unitree’s SDK supports not only inverse kinematics but also simulation-based training. Despite this versatility, much of the current research emphasis has concentrated on developing foundation models for humanoid movement generation, prioritizing broad motor skills and generalization. While such efforts are important, they have left other multi-purpose applications, such as embodied haptic feedback, comparatively underexplored. These resources are still in an early stage, yet they indicate a rapid improvement in the near future. This suggests that current shortcomings may be temporary rather than fundamental, making it timely to investigate humanoids as emerging haptic media.

\subsubsection{Future Directions}
Looking ahead, humanoid-mediated haptics offer promising opportunities across diverse domains. In training and education, repeatable whole-body cues could reinforce skill acquisition and retention. In rehabilitation and healthcare, adaptive stimulation may assist motor recovery and support therapeutic exercises. In entertainment and immersive storytelling, embodied motion feedback could create experiences unattainable through conventional haptic devices.

The versatility of humanoids is well-suited to combining passive and active haptic roles. Passive feedback occurs when the robot responds to user-initiated contact, such as providing resistance when the user pushes on a virtual wall or adjusting its apparent stiffness through a shared physical intermediary (e.g., a bar or plate) to suggest different materials or loads. Similar impedance-based modulation could change the perceived stiffness of a virtual steering wheel or joystick as road conditions or system state vary, while the humanoid grasps the same physical prop the user is holding. Active feedback, in contrast, occurs when the robot initiates contact, such as delivering punch-like impacts or simulating objects that fly toward and collide with the user’s body during a VR boxing or combat game. This dual capability supports both passive, user-initiated contact and active, robot-initiated contact in a controlled, embodied way, which is difficult to achieve with fixed platforms or wearable devices alone.

Because humanoids share the basic morphology and scale of the human body, they are naturally suited to human-centric environments. They can stand behind a chair, approach from the side of a desk, or move between pieces of furniture while using their hands or arms as haptic surfaces, or by interacting with everyday objects already present in the scene. This form factor allows many of the above scenarios to be realized with minimal additional fixtures or custom-built hardware.

Beyond these domain applications, several research directions emerge. With their high degrees of freedom, humanoids can replicate large-scale encountered-type haptics or multi-user collaboration scenarios that were previously enabled through human surrogates or hidden actuators (e.g., HapticTurk~\citep{hapticturk}, TurkDeck~\citep{turkdeck}, TightGame~\citep{maekawa2020tight}). On the HRI side, future studies could examine how social interaction with a humanoid alters its effectiveness as a haptic partner—for example, how brief social exchanges before a session, or mid-task adjustments that change the intensity or range of feedback, influence users’ perceived trust, comfort, or realism. Adaptive feedback based on real-time sensing of user state also presents opportunities. During a VR boxing game, for instance, the humanoid could sense the user’s affective or physical state and dynamically adjust the difficulty, or detect when the user is struggling with VR usage and provide context-aware assistance. Finally, addressing trust and social perception remains a key challenge. As prior work suggests, trust in HRI is fundamentally linked to a user's willingness to be vulnerable to the robot's actions \cite{yagoda2012you}. In our study, several participants reported feeling anxiety during the interaction. Qualitative feedback revealed that this stemmed from a lack of trust in the humanoid's capabilities (P1, P2, P4, P16) or discomfort with its visual appearance (P9, P12). Conversely, participants who did not experience anxiety attributed their sense of safety to a general faith in machines (P14) or explicitly mentioned that the chair served as a medium that mitigated their anxiety (P5, P7, P11, P13).

Realizing these opportunities will require advances on both technical and experiential fronts. Hardware limitations, such as latency, power, and safety, must be overcome, while user acceptance depends on careful experience-centered design that accounts for discomfort and the social perception of robotic motion. Our exploratory study highlights humanoids not merely as social or assistive agents but as potential haptic media. Establishing them as a new modality of VR haptics will require both technological advancements and design strategies that seamlessly integrate human-scale feedback into everyday immersive experiences.

\subsection{Limitations: Scope of Study}
Our exploration was limited in several respects. First, we focused exclusively on a driving simulation with short one-minute sessions, underestimating issues of long-term comfort and fatigue. Although we used a Latin square design and mandatory breaks, the potential delay of simulator sickness between conditions could not be completely ruled out, given the slow decay of motion sickness symptoms \citep{keshavarz2011validating}. Second, the relatively small sample size (n=16) restricts statistical power, and the \textcolor{human}{\textit{Human+Controller}} condition lacked strict standardization, making direct comparisons less controlled. Third, practical constraints such as robot overheating, torque loss, and operational noise further narrowed the scope of evaluation. Fourth, we also did not directly measure the forces delivered by the humanoid or the human; rather, we focused on participants’ subjective evaluations. Additionally, our evaluation of simulation sickness lacked a direct comparison with a standard commercial motion simulator baseline. While we established a visual-only baseline (No-Feedback), future work should benchmark humanoid-induced sickness against established motion platforms to fully contextualize these findings. These factors caution against broad generalization. Extending the study to longer sessions, larger and more diverse participant groups, and additional VR tasks will be essential to validate the broader applicability of humanoid-mediated haptics. 

\section{Conclusion}
We introduced HumanoidTurk, a proof-of-concept system that repurposes a general-purpose humanoid robot as a provider of VR haptics. Moving beyond the conventional view of humanoids as social or assistive agents, our work explored their potential as embodied haptic media. Through a synthesis pipeline mapping in-game g-forces to robot-driven chair motion feedback, we compared humanoid-mediated feedback against traditional modalities. Our user study showed that while humanoid feedback imposed slight costs in comfort and sickness, it substantially enhanced immersion, realism, and enjoyment. Interviews further revealed distinctive opportunities and challenges, highlighting the uniqueness of humanoid-mediated haptics as a human-scale modality. This study thus contributes an early step toward positioning humanoids not only as interactive partners but also as versatile haptic media, opening new design spaces for immersive experiences.

\begin{acks}
This research was financially supported by the Ministry of Trade, Industry, and Energy (MOTIE), Korea, under the “Global Industrial Technology Cooperation Center program” supervised by the Korea Institute for Advancement of Technology (KIAT) (Grant No. P0028435) (Contribution Rate: 80\%). This work was also supported by Institute of Information \& communications Technology Planning \& Evaluation (IITP) grant funded by the Korea government (MSIT) (No.2019–0–01842, Artificial Intelligence Graduate School Program (GIST)) (Contribution Rate: 10\%). This work was also supported by GIST-IREF from Gwangju Institute of Science and Technology (GIST) (Contribution Rate: 5\%). This research was partially funded by the JST PRESTO Grant Number JPMJPR23I5 (Contribution Rate:5\%).
\end{acks}

\bibliographystyle{ACM-Reference-Format}
\bibliography{sample-base}
\balance

\onecolumn
\newpage
\appendix

\section{Haptic Feedback Synthesis Details}
While the Method section introduced our two synthesis approaches (filter-based vs. threshold-based), here we provide implementation-level details not included in the main text.

\subsection{Smoothed (filter-based) synthesis}
We employed simple one-pole filters for real-time decomposition:
\begin{itemize}
    \item Low-pass cutoff: $0.7$ Hz (tilt/sway, accel guard trend).
    \item High-pass cutoff: $0.4$ Hz (for jolt detection).
    \item Roughness RMS window: $0.2$ s (for vibration).
\end{itemize}

Longitudinal cues (gy) were combined with the following rules:
\begin{itemize}
    \item \textit{Tilt}: LPF(gy) $\rightarrow$ continuous pitch sway (gain = 0.1 m/g).
    \item \textit{Jolt}: HPF(gy) $\rightarrow$ brief transient cue (gain = 0.5 m/g).
    \item \textit{Vibration}: RMS(HPF) $\rightarrow$ 12 Hz sinusoidal vibration (amplitude gain = 0.3 m).
    \item During jolt periods, tilt and vibration weights were reduced to highlight the transient.
\end{itemize}

\subsection{Threshold-based synthesis}
For comparison, the baseline mapping applied minimal filtering:
\begin{itemize}
    \item Low-pass cutoff: $0.7$ Hz (sway, accel guard trend).
    \item Deadzone: 0.2 g.
    \item Clip range: $\pm 1.3$ g (force), $\pm 5.0$ g/s (jerk).
    \item Gain: 0.1 m/g for gy, 0.1 m/(g/s) for jerk.
\end{itemize}

\subsection{Additional General Rules}
To avoid contradictory impulses during sustained acceleration:
\begin{itemize}
    \item Enter acceleration mode if LPF(gy) $> 0.15$ g.
    \item Release if LPF(gy) $< 0.10$ g.
    \item Minimum hold time: 0.20 s.
    \item Opposite-direction jolts suppressed (default: 100\%).
\end{itemize}

\textbf{Virtual Rod.}
All rotations were applied to a ``virtual rod'' connecting the robot’s two hands. The rod’s midpoint served as the pivot, with yaw rotation added before inverse kinematics.

\newpage
\section{User study demographics}
The following table presents the demographics.
\begin{table}[H]
\centering
\caption{Demographic Information of Participants.}
\label{tab:demographics}
\Description{Table showing demographic information of 16 participants. Each row lists participant ID, age, gender, height, weight, driving experience (years), simulator experience, and VR experience. Ages range from 19 to 31 years; 7 participants are female and 9 are male. Heights range from 158 cm to 184 cm, and weights from 48 kg to 76 kg. Most had little or no driving experience, with only a few reporting up to 4 years. Simulator experience ranged from 0 to 10, and VR experience ranged from 1 to 6. Overall, the participant group consisted of young adults with varied prior simulator and VR exposure but limited driving experience.}
\begin{tabular}{cccccccc}
\toprule
Participant & Age & Gender & Height (cm) & Weight (kg) & Driving Exp. (yr) & Simulator Exp. & VR Exp. \\
\midrule
         P1 &  23 &        F&       165.0 &        60.0 &                 0 &             10 &       4 \\
         P2 &  24 &        F&       163.0 &        50.0 &                 0 &              3 &       4 \\
         P3 &  20 &        F&       158.0 &        48.0 &                 0 &              0 &       3 \\
         P4 &  19 &        M&       170.0 &        56.0 &                 0 &              0 &       1 \\
         P5 &  20 &        M&       170.0 &        57.0 &                 1 &              0 &       2 \\
         P6 &  19 &        F&       162.0 &        50.0 &                 0 &              0 &       5 \\
         P7 &  24 &        M&       183.0 &        76.0 &                 0 &              0 &       5 \\
         P8 &  23 &        M&       173.0 &        75.0 &                 0 &              0 &       4 \\
         P9 &  22 &        F&       162.0 &        50.0 &                 2 &              1 &       2 \\
        P10 &  23 &        M&       182.0 &        76.0 &                 4 &              1 &       4 \\
        P11 &  23 &        M&       174.0 &        76.0 &                 0 &              1 &       2 \\
        P12 &  26 &        M&       177.0 &        74.0 &                 0 &              0 &       3 \\
        P13 &  23 &        M&       178.0 &        72.5 &                 0 &              0 &       3 \\
        P14 &  31 &        F&       162.7 &        50.0 &                 0 &              0 &       4 \\
        P15 &  22 &        M&       184.0 &        71.0 &                 4 &              0 &       2 \\
        P16 &  24 &        M&       165.0 &        64.0 &                 0 &              4 &       5 \\
\bottomrule
\end{tabular}
\end{table}

\section{Additional Questionnaires}
These five items (immersion, realism, comfort, enjoyment, suitability) were self-authored single-item questions.
\begin{itemize}
    \item \textbf{Immersion}: When playing with this feedback, how immersed did you feel? (1 = Not immersed at all, 7 = Very immersed)
    \item \textbf{Realism}: When playing with this feedback, how realistic did the experience feel? (1 = Not realistic at all, 7 = Very similar to reality)
    \item \textbf{Comfort}: When playing with this feedback, how comfortable was the experience? (1 = Very uncomfortable, 7 = Very comfortable)
    \item \textbf{Enjoyment}: When playing with this feedback, how enjoyable was the experience? (1 = Not enjoyable at all, 7 = Very enjoyable)
    \item \textbf{Suitability}: How suitable was this feedback for the game? (1 = Not suitable at all, 7 = Very suitable)
\end{itemize}

\section{UEQ-S Detailed Results}
\begin{table}[!h]
  \centering
  \caption{Mean and standard deviation (SD) of PQ and HQ scores (-3 to +3 scale, Mean (SD))}
  \label{tab:PQHQ}
\Description{Table summarizing mean and standard deviation (SD) of Pragmatic Quality (PQ) and Hedonic Quality (HQ) scores on a –3 to +3 scale across four conditions. For No-feedback: PQ -0.47 (0.97), HQ -1.23 (1.08). For Controller: PQ 0.42 (0.93), HQ –0.11 (1.21). For Controller+Humanoid: PQ 0.70 (1.44), HQ 1.88 (1.05). For Controller+Human: PQ 0.84 (1.16), HQ 1.22 (0.89). Overall, humanoid and human conditions scored higher on both PQ and HQ, with the humanoid condition showing the highest HQ ratings.}
  \begin{tabular}{lcc}
    \toprule
    \textbf{Condition} & {\textbf{Pragmatic Quality}} & {\textbf{Hedonic Quality}} \\
    \midrule
    No-Feedback           & -0.47 (0.97) & -1.23 (1.08) \\
    Controller            &  0.42 (0.93) & -0.11 (1.21) \\
    Humanoid+Controller &  0.70 (1.44) &  \textbf{1.88 (1.05)} \\
    Human+Controller    &  \textbf{0.84 (1.16)} &  1.22 (0.89) \\
    \bottomrule
  \end{tabular}
\end{table}

\begin{table*}[!ht]
\centering
\caption{Mean and standard deviation of UEQ-S items by condition (-3 to 3 scale, Mean (SD))}
\label{tab:ueqs_item_scores}
\Description{Table presenting mean and standard deviation (SD) of UEQ-S item scores (-3 to +3 scale) across four conditions: No-feedback, Controller, Controller+Humanoid, and Controller+Human. Items include Supportiveness, Simplicity, Efficiency, Clarity, Excitement, Interest, Inventiveness, and Novelty. Supportiveness: highest for Controller+Humanoid (0.75), positive in other conditions. Simplicity: rated highest in Controller+Human (1.75), followed by Controller+Humanoid (1.06). Efficiency: negative in No-feedback and Controller, positive in humanoid and human conditions. Clarity: negative in No-feedback, positive in others, highest for Controller+Humanoid (0.69). Excitement: strongly positive for Controller+Humanoid (2.12) and Controller+Human (1.62). Interest: highest for Controller+Humanoid (2.00), followed by Controller+Human (1.62). Inventiveness: highest in Controller+Humanoid (1.50), positive in Controller+Human (0.75), negative in others. Novelty: lowest in No-feedback (–2.00), highest in Controller+Humanoid (1.88). Overall, humanoid feedback (Controller+Humanoid) consistently scored highest across most dimensions, particularly Excitement, Interest, Inventiveness, and Novelty.}
\begin{tabular}{lcccc}
\toprule
\textbf{Item} & \textbf{No-Feedback} & \textbf{Controller} & \textbf{Humanoid+Controller} & \textbf{Human+Controller} \\
\midrule
Supportiveness & -0.50 (1.26) & 0.62 (0.72) & \textbf{0.75 (1.53)} & 0.62 (1.26) \\
Simplicity & 0.00 (1.41) & 1.00 (1.26) & 1.06 (1.84) & \textbf{1.75 (1.06)} \\
Efficiency & -0.75 (1.29) & -0.38 (1.36) & 0.31 (1.66) & \textbf{0.44 (1.41)} \\
Clarity & -0.62 (1.93) & 0.44 (1.46) & \textbf{0.69 (1.74)} & 0.56 (1.79) \\
Excitement & -0.62 (1.50) & 0.38 (1.59) & \textbf{2.12 (0.89)} & 1.62 (1.02) \\
Interest & -0.38 (1.63) & 0.94 (1.29) & \textbf{2.00 (1.15)} & 1.62 (1.15) \\
Inventiveness & -1.94 (1.24) & -0.81 (1.38) & \textbf{1.50 (1.41)} & 0.75 (1.34) \\
Novelty & -2.00 (1.03) & -0.94 (1.34) & \textbf{1.88 (1.31)} & 0.88 (1.50) \\
\bottomrule
\end{tabular}
\end{table*}

\newpage
\section{SSQ Detailed Results}
\begin{table*}[!h]
\centering
\caption{Mean and standard deviation of individual items of SSQ (0 to 4 scale, Mean (SD))}
\label{tab:ssq_all_items}
\Description{Table showing mean (SD) scores of individual simulator sickness symptoms across four conditions: No-feedback, Controller, Controller+Humanoid, and Controller+Human. General discomfort: lowest in Controller (0.25), highest in Controller+Humanoid (1.00). Fatigue: minimal in No-feedback (0.12), higher in humanoid (0.56). Headache: low across all conditions (~0.12). Eyestrain: around 0.31 in No-feedback and Controller, slightly lower in humanoid (0.25). Difficulty focusing: highest in humanoid (0.31), minimal in Controller (0.06). Nausea: low overall, 0.12 in humanoid and human. Difficulty concentrating: highest in humanoid (0.50). Blurred vision: most frequent in No-feedback (0.19). Vertigo: around 0.12 in most conditions. Other symptoms (salivation, sweating, dizziness, stomach awareness, burping) remained near zero. Overall, most symptoms were low across all conditions, but Controller+Humanoid showed relatively higher ratings for general discomfort, fatigue, and difficulty concentrating.}
\begin{tabular}{lcccc}
\toprule
\textbf{Symptom} & \textbf{No-Feedback} & \textbf{Controller} & \textbf{Humanoid+Controller} & \textbf{Human+Controller} \\
\midrule
General discomfort & 0.31 (0.48) & 0.25 (0.45) & \textbf{1.00 (0.89)} & 0.44 (0.51) \\
Fatigue & 0.12 (0.34) & 0.19 (0.40) & \textbf{0.56 (0.63)} & 0.38 (0.62) \\
Headache & \textbf{0.12 (0.34)} & 0.06 (0.25) & \textbf{0.12 (0.34)} & 0.06 (0.25) \\
Eyestrain & \textbf{0.31 (0.48)} & \textbf{0.31 (0.48)} & 0.25 (0.45) & 0.19 (0.40) \\
Difficulty focusing & 0.19 (0.54) & 0.06 (0.25) & \textbf{0.31 (0.60)} & 0.19 (0.40) \\
Increased salivation & 0.00 (0.00) & \textbf{0.06 (0.25)} & 0.00 (0.00) & 0.00 (0.00) \\
Sweating & 0.00 (0.00) & 0.06 (0.25) & \textbf{0.19 (0.40)} & 0.06 (0.25) \\
Nausea & 0.06 (0.25) & 0.06 (0.25) & \textbf{0.12 (0.50)} & \textbf{0.12 (0.34)} \\
Difficulty concentrating & 0.12 (0.34) & 0.06 (0.25) & \textbf{0.50 (0.73)} & 0.25 (0.45) \\
Fullness of head & \textbf{0.12 (0.34)} & \textbf{0.12 (0.34)} & \textbf{0.12 (0.34)} & 0.06 (0.25) \\
Blurred vision & \textbf{0.19 (0.40)} & 0.00 (0.00) & 0.12 (0.34) & 0.00 (0.00) \\
Dizzy eyes open & 0.00 (0.00) & 0.06 (0.25) & \textbf{0.12 (0.34)} & 0.06 (0.25) \\
Dizzy eyes closed & 0.12 (0.34) & 0.06 (0.25) & \textbf{0.19 (0.40)} & 0.06 (0.25) \\
Vertigo & \textbf{0.12 (0.34)} & \textbf{0.12 (0.34)} & \textbf{0.12 (0.34)} & 0.06 (0.25) \\
Stomach awareness & 0.00 (0.00) & 0.00 (0.00) & \textbf{0.06 (0.25)} & \textbf{0.06 (0.25)} \\
Burping & 0.00 (0.00) & 0.00 (0.00) & 0.00 (0.00) & 0.00 (0.00) \\
\bottomrule
\end{tabular}
\end{table*}

\end{document}